# A New Method to Constrain the Appearance and Disappearance of Observed Jellyfish Galaxy Tails

Rory Smith,[1] Jong-Ho Shinn,[2] Stephanie Tonnesen,[3] Paula Calderón-Castillo,[4] Jacob Crossett,[5]
Yara L. Jaffé,[5] Ian Roberts,[6] Sean McGee,[7] Koshy George,[8] Benedetta Vulcani,[9] Marco Gullieuszik,[9]
Alessia Moretti,[9] Bianca M. Poggianti,[9] and Jihye Shin[2]

[1]*Departamento de Física, Universidad Técnica Federico Santa María, Vicuña Mackenna 3939, San Joaquín, Santiago de Chile*
[2]*Korea Astronomy and Space Science Institute (KASI), 776 Daedeokdae-ro, Yuseong-gu, Daejeon 34055, Korea*
[3]*Center for Computational Astrophysics, Flatiron Institute, 162 5th Ave, New York, NY 10010, USA*
[4]*Departamento de Astronomía, Universidad de Concepción, Casilla 160-C, Concepción, Chile*
[5]*Instituto de Física y Astronomía, Universidad de Valparaíso, Avda. Gran Bretaña 1111 Valparaíso, Chile*
[6]*Leiden Observatory, Leiden University, PO Box 9513, 2300 RA Leiden, The Netherlands*
[7]*University of Birmingham School of Physics and Astronomy, Edgbaston, Birmingham, UK*
[8]*Faculty of Physics, Ludwig-Maximilians-Universität, Scheinerstr 1, Munich, 81679, Germany*
[9]*INAF–Astronomical Observatory of Padova, Vicolo dell'Osservatorio 5, I-35122 Padova, Italy*

## Abstract

We present a new approach to observationally constrain where the tails of Jellyfish (JF) galaxies in groups and clusters first appear and how long they remain visible with respect to the moment of their orbital pericenter. This is accomplished by measuring the distribution of their tail directions with respect to their host's center, and their distribution in a projected velocity-radius phase-diagram. We then model these observed distributions using a fast and flexible approach where JF tails are painted onto dark matter halos according to a simple parameterised prescription, and perform a Bayesian analysis to estimate the parameters. We demonstrate the effectiveness of our approach using observational mocks, and then apply it to a known observational sample of 106 JF galaxies with radio continuum tails located inside 68 hosts such as groups and clusters. We find that, typically, the radio continuum tails become visible on first infall when the galaxy reaches roughly three quarters of $r_{200}$, and the tails remain visible for a few hundred Myr after pericenter passage. Lower mass galaxies in more massive hosts tend to form visible tails further out and their tails disappear more quickly after pericenter. We argue that this indicates they are more sensitive to ram pressure stripping. With upcoming large area surveys of JF galaxies in progress, this is a promising new method to constrain the environmental conditions in which visible JF tails exist.

*Keywords:* Galaxy clusters (584) – Galaxy dark matter halos (1880) – Galaxy evolution (594) – Galaxy groups (597) – Galaxy tails (2125) – Orbits (1184) – Radio continuum emission (1340)

## 1. INTRODUCTION

For several decades, it has been recognised that the galaxies become increasingly deficient in atomic gas (HI) in dense environments such as clusters or massive groups (Davies & Lewis 1973; Huchtmeier et al. 1976; Haynes & Giovanelli 1986; Solanes et al. 2001; Boselli

Corresponding author: Rory Smith
rorysmith274@gmail.com

& Gavazzi 2006), compared to their counterparts in the field (Chamaraux et al. 1980; Solanes et al. 1996; Toribio et al. 2011; Dénes et al. 2014; Jones et al. 2018). The loss of HI gas is believed to be of significance for galaxy evolution (see Cortese et al. 2021 for a recent review), and a net reduction in galaxy star formation rates is observed in HI deficient galaxies (Kennicutt 1983; Gavazzi et al. 2002, 2006, 2013). The reduction in star formation tends to preferentially occur from the outside inwards, resulting in truncated star forming disks (e.g., Koopmann & Kenney 2004a,b; Koopmann et al. 2006; Rose et al. 2010; Cortese et al. 2012; Fossati et al. 2013; Gullieuszik et al. 2017; Finn et al. 2018). If all of the gas content of a galaxy is removed, it will be fully quenched. One probable candidate for the cause of the HI deficiency is ram pressure stripping (Gunn & Gott 1972), perhaps in combination with viscous stripping (Nulsen 1982) and thermal evaporation (Cowie & Songaila 1977). As a disk galaxy passes through a cluster on its orbit, its disk gas is subjected to ram pressure as it pushes through the hot intracluster medium (ICM) within the cluster. This pressure preferentially strips away disk gas from the outer disk where it is more easily unbound, creating an outside-in HI gas disk truncation that is observed (e.g., Warmels 1988; Cayatte et al. 1990; Bravo-Alfaro et al. 2000, 2001; Chung et al. 2009a; Scott et al. 2010). Although other non-hydrodynamical environmental mechanisms (e.g., gravitational tides; Valluri & Jog 1990, 1991; Moore et al. 1996, 1999; Gnedin 2003; Smith et al. 2015) can potentially strip gas from the outer disks of galaxies, hydrodynamical stripping mechanisms can strip the disk gas deep inside the optical disk of the galaxy, while leaving the stellar disk dynamically undisturbed (Vollmer et al. 2001a; Vollmer 2003; Vollmer et al. 2004) (except in gas rich galaxies where the removal of the gas potential may perturb the stellar disk, see Smith et al. 2013).

In nearby clusters, it is possible to catch the gas stripping process in action. Some gas deficient spirals present prominent long, one-sided HI tails that emanate from truncated gas disks and are spatially resolved in 21cm maps thanks to their proximity (e.g., Oosterloo & van Gorkom 1975; Chung et al. 2009b; Ramatsoku et al. 2019; Deb et al. 2020a). In some cases, extragalactic star formation is observed to occur within the stripped gas streams (e.g., Owen et al. 2006; Cortese et al. 2007; Sun et al. 2007; Yoshida et al. 2008; Smith et al. 2010; Hester 2010; Owers et al. 2012; Fumagalli et al. 2011; Ebeling et al. 2014; McPartland et al. 2016; Poggianti et al. 2016a; George et al. 2018). Such galaxies tend to have moderately enhanced star formation rate in their disks as well (e.g., Vulcani et al. 2018a, 2020; Roberts & Parker 2020a). The presence of external star formation originally earned such galaxies the title of 'Jellyfish' (JF) galaxies, as the streams of luminous young stars gave the impression of JF tentacles. But more recently, the term JF galaxy has been used less strictly, to describe any with long one-sided streams of stripped material whether it is star forming or not and, in this study, we follow this latter naming convention.

Recently, there has been a boom in interest in JF galaxies (for a recent review see Boselli et al. 2022). The streams of ram pressure stripped gas or extragalactic star formation have been revealed and studied using multiple different tracers in nearby clusters and groups. Each tracer provides a diferent window on the gas properties within the JF tails. $H_\alpha$ imaging relates to ionised gas at temperatures of roughly $10^4$ K (Yagi et al. 2017; Fossati et al. 2018; Gavazzi et al. 2018). The VESTIGE survey (Boselli et al. 2018), which is a deep blind $H_\alpha$ survey in the Virgo cluster, is a good recent example of this. IFU observations by the 'GAs Stripping Phenomena in galaxies with MUSE' (GASP, Poggianti et al. 2017) survey have produced numerous detailed views of the ionised gas and star formation in JF galaxy tails e.g., Gullieuszik et al. 2017; Vulcani et al. 2018b; Poggianti et al. 2019a; Bellhouse et al. 2021. In star-forming galaxies, radio continuum observations are sensitive to synchotron radiation from cosmic rays accelerated by supernovae. While undergoing ram pressure, tails of synchotron emission may be observed as the cosmic rays are stripped from the galaxy disk (Botteon et al. 2020; Chen et al. 2020; Lal 2020; Roberts et al. 2021a,b, 2022). 21cm observations can trace out the neutral atomic gas in JF tails (Minchin et al. 2019; Deb et al. 2020b; Healy et al. 2021; Wang et al. 2021; Loni et al. 2021). Meanwhile, X-ray observations can trace out very hot ionised gas in JF tails (Boselli et al. 2016; Wood et al. 2017; Ge et al. 2019; Poggianti et al. 2019b; Campitiello et al. 2021; Sun et al. 2021). Some JF galaxies were observed with multiple high resolution observations (such as optical imaging with the Hubble-space telescope and sub-mm observations with ALMA) which has enabled a detailed study of the gas properties in their tails (Abramson et al. 2016; Jáchym et al. 2017; Cramer et al. 2019, 2021).

Many JF galaxies often appear to be undergoing a dramatic transformation, some with ionised gas tails extending nearly 100 kpc in length (e.g., Yagi et al. 2010; Smith et al. 2010; Bellhouse et al. 2019). But we lack a clear understanding of the conditions in which the tails can form and survive. Even in the simplified case of a fixed ram pressure, we can expect that lower density gas would be more rapidly accelerated from the disk, followed later by more dense gas (Tonnesen & Bryan 2009), meaning tails may appear different depending on the observational tracer that is used. Furthermore, the physics within the streams of stripped material is complex, and poorly understood. The tail gas finds itself bathed in the hot ICM, and a combination of turbulence, magnetic fields, and feedback from newly formed stars can alter the tail gas properties and eventually mix out the tail gas into the ICM (Vollmer et al. 2001a; Tonnesen & Stone 2014; Vijayaraghavan & Sarazin 2017; Müller et al. 2020; Gronke et al. 2022; Sun et al. 2021). The amount of material that is stripped evolves with the changing ram pressure, and the gas in the tails can have a wide range of densities and velocities.

The complexity of the physics involved means it is difficult to model the tails accurately or predict the conditions under which JF tails will first become visible, or how long they will remain visible. But a better understanding of this could potentially provide valuable clues on the gas stripping and quenching process that newly accreted galaxies are subjected to on entering dense environments. The distance at which the tails appear from the cluster center can show when ram pressure first becomes effective, and the quenching process begins. This may be initiated when galaxies first enter the cluster, or it may occur much later if galaxies are initially surrounded by inflowing streams of ambient gas that infall alongside them (Bahé et al. 2013). Some models predict ram pressure stripping could begin as far out as ~2-3 virial radii due to accretion shocks (Zinger et al. 2018). The length of time that their tails remain visible may also provide insights on the survival of disk gas to ram pressure. For example, if galaxies are fully stripped of their gas on first infall, then their tails may disappear prior to pericenter passage.

To try to improve our understanding of these issues in a cosmological context, a number of studies have searched for JF galaxies in large volume cosmological hydrodynamical simulations that contain cluster mass objects within their volume (Bahé et al. 2013; Bahé & McCarthy 2015; Marasco et al. 2016; Jung et al. 2018; Yun et al. 2019; Lotz et al. 2019; Troncoso-Iribarren et al. 2020). However, it is uncertain if all of the important physics required to model the stripping process are included or treated sufficiently accurately. In many cases, massive clusters tend to overquench their satellites (Brown et al. 2017; Davé et al. 2017; Stevens et al. 2019; Xie et al. 2020; Donnari et al. 2020), perhaps as a result of insufficient resolution and/or an interaction between sub-grid physics feedback implementations and the ram pressure stripping process (Bahé & McCarthy 2015; Emerick et al. 2016; Kazantzidis et al. 2017; Zoldan et al. 2017; De Lucia et al. 2019). Therefore, it is not clear if the conditions under which JF tails appear and how long they last in these simulations can be assumed to directly match the observed objects.

As an alternative to self-consistent hydrodynamical cosmological simulations, some studies have had success applying analytical models to observed ram pressure stripped galaxies and their tails (Jaffé et al. 2015, 2018; Gullieuszik et al. 2020). In this case, the strength of the ram pressure is estimated from a galaxy's location in a projected velocity-radius phase-space (PS) diagram, combined with an estimate for the ICM radial density distribution. There are already some uncertainties introduced as projection effects may affect the galaxy's location in phase-space, and, even if X-ray observations are available for the cluster of interest, they are normally only available near the cluster center and extrapolated out to larger clustocentric radii. The impact of this estimated ram pressure on the disk galaxy is then calculated using an analytical equation based on a Gunn & Gott (1972) formalism, where the ram pressure is compared to the disk restoring force at different radii throughout the disk. As a result there are many factors that can affect the stripping efficiency including the shape and scalelength of the radial disk density profile, how extended the gas disk is with respect to the stars (Cayatte et al. 1994), and the restoring force from a stellar bulge and the dark matter halo (Abadi et al. 1999), and it is challenging to know all of these properties for a large sample, especially given we would really need to know them prior to the onset of stripping. The gas disk truncation radius is also calculated assuming face-on stripping, although hydrodynamical simulations show this assumption is reasonable to first order, as long as the disk is not close to edge-on to the ram pressure wind (Vollmer et al. 2001b; Roediger & Brüggen 2007).

Given the current limitations of cosmological hydrodynamical simulations, and the large number of parameters and uncertainties in the existing analytical models, in this study we seek to develop a novel approach that has less parameter dependencies. The aim is to try to better determine the conditions under which the tails of observed JF galaxies appear and when they disappear. This new approach combines two observational measurements; the distribution of JF tail angles measured with respect to their group/cluster center, and their observed locations in a projected velocity-radius phase diagrams.

As larger samples of JF galaxies have been accumulated, the distributions of their tail directions have been increasingly used to interpret whether galaxies are on first infall into their hosts. This has been attempted observationally in HI (Chung et al. 2007), optical (Poggianti et al. 2016b; Roberts & Parker 2020b), H-alpha (Liu et al. 2021; Durret et al. 2021), UV (Smith et al. 2010), X-ray (Merrifield 1998) and radio continuum (Roberts et al. 2021b,a). The general conclusion from these studies is that the high fraction of objects with tails pointing away from their host center implies these objects are on their first infall into their host along quite radial orbits. Unlike the observations listed above, a much flatter tail angle distribution was measured in the hydrodynamical cosmological simulations of Yun et al. (2019).

Projected velocity-radius phase-space diagrams have also been shown to provide useful constraints on the time since galaxies first fell into their hosts (Muzzin et al. 2014; Oman & Hudson 2016; Rhee et al. 2017a; Pasquali et al. 2019). A number of authors have applied these diagrams to samples of JF galaxies (McPartland et al. 2016; Yoon et al. 2017; Jaffé et al. 2018; Roberts & Parker 2020b; Roberts et al. 2021b,a; Liu et al. 2021). Similar to the results based on tail directions, the general conclusion drawn is that JF galaxies tend to prefer the 'recent-infaller' area of the phase-space diagram (see for example Figure 6 of Rhee et al. 2017b).

However, this study represents the first attempt to simultaneously use the observed JF tail directions combined with their location in phase-space in a complementary manner to try to quantitatively constrain where their tails form and how long they last. This is accomplished using cosmological simulation results and Bayesian parameter estimation. We apply our novel method to a sample of LOFAR radio continuum JF galaxies in groups and clusters. The content of our paper can be summarised as follows. Our observational sample is described in Section 2.1, and our cosmological simulations are described in Section 2.2. Our method is described in Section 3, and we test the method with mock observational data sets in Section 4. The results of applying the method to the full JF sample are presented in Section 5.1, and to JF subsamples in Section 5.2. We summarise and conclude in Section 6. Throughout this study, we assume a $\Lambda$CDM cosmology of $\Omega_m = 0.3$, $\Omega_\Lambda = 0.7$, $\Omega_b = 0.047$ and $h_0 = 0.7$.

## 2. SAMPLE

### 2.1. LOFAR Jellyfish in Group/Cluster Mass Hosts

Our main observational sample is the sample of Jelly-Fish (JF) galaxies presented in Roberts et al. (2021b,a) (herein R21ab). JF galaxies were visually identified by the presence of 144 Mhz radio continuum tails in groups and clusters covered by the LOFAR two meter sky survey (LoTSS; Shimwell et al. 2017, 2019). Contamination by AGN was found to be less than 5-10% in the cluster sample of R21a. In star-forming galaxies, LoTSS is sensitive to synchrotron emission from cosmic rays accelerated by supernovae. For star-forming galaxies experiencing RPS, tails of synchrotron emission may be observed as these cosmic rays are stripped from the galaxy disk. Thus the visibility of the tails at 144 Mhz requires star formation in the disk, but not necessarily in the tails themselves. The LoTSS survey combines high resolution ($\sim$6") with high sensitivity uniform coverage ($\sim$100 $\mu$Jy/beam) over a large area, and will eventually cover the entire northern sky. The groups and clusters in R21ab are selected from spectroscopic SDSS group/cluster catalogues (Wang et al. 2015; Lim et al. 2017). For simplicity we herein refer to these groups and clusters collectively as 'hosts'. Objects at a projected radius of <1 $R_{180}$ (where $R_{180}$ is the radius containing a region with 180 times the critical density of the Universe) and a line-of-sight velocity difference $< 3\sigma_{rms}$ (where $\sigma_{rms}$ is the velocity dispersion of the host) were considered to be associated with the hosts. All objects have optical counterparts. Galaxy star formation rates (SFR) and stellar masses were taken from the GSWLC-2 catalogue (Salim et al. 2016, 2018) and measured based on SED fitting with CIGALE (Boquien et al. 2019). Only galaxies with a specific star formation rate sSFR> $10^{-11}$ yr$^{-1}$ (i.e., star forming) were included in the sample. Due to the decreasing sensitivity of LoTSS to low-mass, star-forming galaxies (R21a) we also remove galaxies with masses below $10^{9.5}$ M$_\odot$. This suppresses a tendency for us to detect galaxies with lower masses than this limit only for low redshifts which would otherwise make comparison with the simulations challenging.

JF galaxies displaying long tails are expected to be quite asymmetrical in shape. Therefore, we also consider a shape asymmetry parameter defined as

$$A_s = \frac{\Sigma \mid X_0 - X_{180} \mid}{2 \times \Sigma \mid X_0 \mid} \qquad (1)$$

where $X_0$ and $X_{180}$ is the complete radio continuum map of an object and its 180° rotation (Pawlik et al. 2016). We only consider objects with $A_s > 0.3$ which reduces the number of galaxies that we must visually inspect by filtering out symmetrical objects. R21a states that this choice of threshold includes $\sim$ 85% of visually identified LoTSS jellyfish galaxies in clusters, while excluding $\sim$ 70% of LoTSS sources in clusters which are not identified as jellyfish.

We measure the noise level in the background surroundings of each individual galaxy, and find values that range from 40-280 $\mu$Jy/beam. To reduce the impact of varying noise levels on our results, we take a cut for galaxies with a noise level greater than 200 $\mu$Jy/beam, which are likely associated with artifacts or bright contaminating sources. The final sample of JF contains 106 galaxies in 68 hosts. We also test a more conservative 150 $\mu$Jy/beam noise-cut but find that this reduces our number statistics to only 89 galaxies, and the results do not change significantly (see the Appendix; Fig. 11).

Histograms of the redshift, stellar mass and host mass distribution of our main sample can be found in Fig. 1. The galaxies have stellar masses in the range $10^{9.5}$–$10^{11}$ M$_\odot$, and their hosts range in mass from low mass hosts ($10^{12.5}$–$10^{13}$ $M_\odot$) to massive clusters ($10^{14.5}$–$10^{15}$ $M_\odot$). According to R21b, the observed fraction of JF is higher in more massive hosts. The redshift range is from z=0.015 to 0.05. The lower limit is because the SDSS group catalogues contain few galaxies with z<0.01. The imposed upper limit is a little arbitrary but with increasing redshift it becomes increasingly difficult to detect low surface brightness tails and short tails become more difficult to resolve. Some examples of JF galaxies (radio continuum contours overlayed on optical imaging) from our sample can be seen in Fig. 2. For similar images of the full sample, see the Appendices in R21ab.

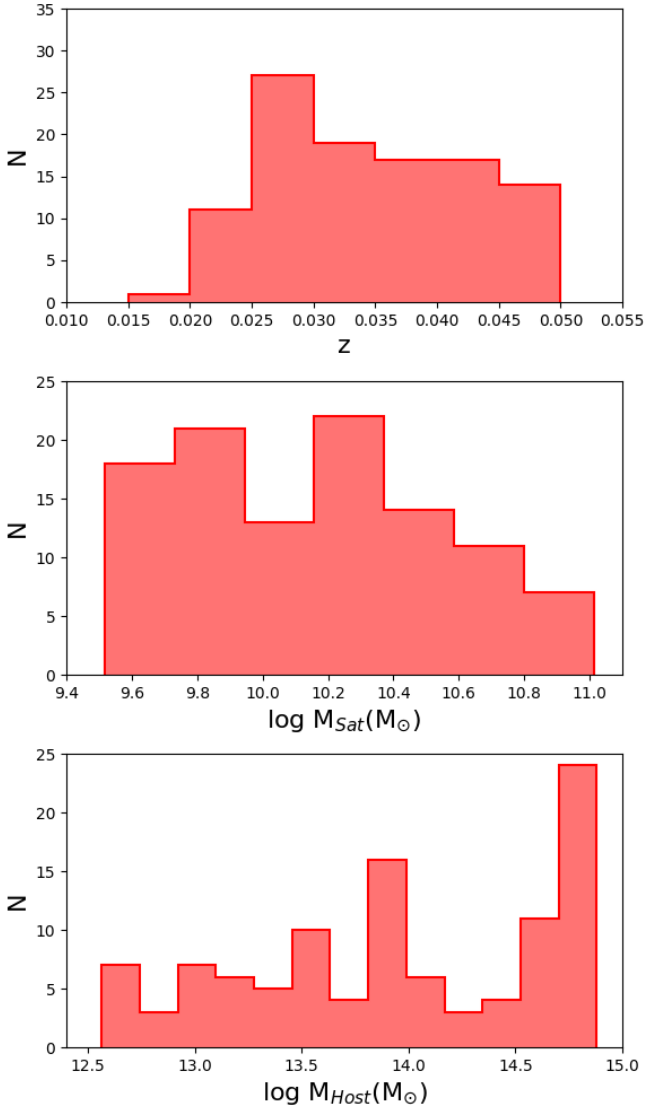

**Figure 1.** General properties of the main LOFAR Jellyfish sample. Top: redshift distribution, center: distribution of stellar masses of satellite galaxies, bottom: distribution of host masses.

### 2.2. N-body Cosmological Simulations

To model the observations, we use an N-body cosmological simulation that was run as a dark matter only model using the GADGET-3 code (Springel et al. 2001). To generate the initial conditions at redshift=200, the Multi Scale Initial Condition software (MUSIC; Hahn & Abel 2011) was used. The CAMB package (Code Anisotropies in the Microwave Background; Lewis et al. 2000) is used to calculate the linear power spectrum. We analyse a single cosmological volume with dimensions $120 \times 120 \times 120$ Mpc/h. The dark matter particles have a fixed mass of $1.072 \times 10^9 M_\odot/h$ in all the simulations. Output files were produced at $\sim$100 Myr intervals consisting of the mass, velocities and positions of dark matter particles for each snapshot, down to redshift zero.

We then run the ROCKSTAR halo finder on the dark matter particle files to build a catalogue of halos at each snapshot (Behroozi et al. 2013). In ROCKSTAR, halos are identified using a hierarchical Friends-of-Friends (FoF) approach that combines six-dimensional phase space information and one-dimensional time, and provides information on the merging history as well. We remove all halos with a mass lower than $10^{11} M_\odot$ in order to ensure our halos are well resolved (more than 100 particles per halo). The merger tree is built with Con-sistent Trees (Behroozi et al. 2013), which combines particle IDs with halo trajectory information to improve the linking of halos between snapshots. In total, there are 692 host halos (central halos with a mass greater than $10^{13}$ $M_\odot$) in our simulation volume, and we record 22813 first infaller halos that fall into these hosts.

### 3. METHOD

#### 3.1. Painting Tails onto First Infaller Satellite Halos in Dark Matter Only Cosmological Simulations

Using the cosmological simulations described in Section 2.2, we produce simulated projected tail angle histograms and simulated locations in projected phase-space for comparison with the observations. As our cosmological simulations are dark matter only, there are no genuine hydrodynamical gas tails within them. Therefore, our approach is to 'paint-on' the tails. We add them on and remove them, according to different assumptions in our model about their visibility. The three main parameters we consider in the model are listed below (also see Fig. 3).

*r1:* For an object on first infall into the host, the JF tail is assumed to first become visible at the 3D radius *r*1, which is given as a percentage of the host's $r_{200}$ (the radius containing a volume with 200 times the critical density of the Universe), and allowed to vary from 10% to 200% in steps of 1%. For example, 200% means the JF tails appear when the galaxy is infalling for the first time and passes the 3D radius of 2 $r_{200}$ from the host center. In Jaffé et al. (2018), optical JF candidates were seen out to projected radii of close to 2 $r_{200}$, and so we provide a similar extended range of $r1$ to allow a wide parameter space to be explored. We note that, although the LOFAR sample is limited to be inside a projected radius of $R_{180}$ (for comparison, $R_{180}$ is typically 5% larger than $r_{200}$; Reiprich et al. 2013), we can still model those galaxies with 3D radii out to 2 $r_{200}$. Thanks to projection effects, some of these objects will fall inside of the limiting projected radius and can be considered in the modelling. Therefore, to account for this, we simply use a projected radius limit on the simulations that matches the observations.

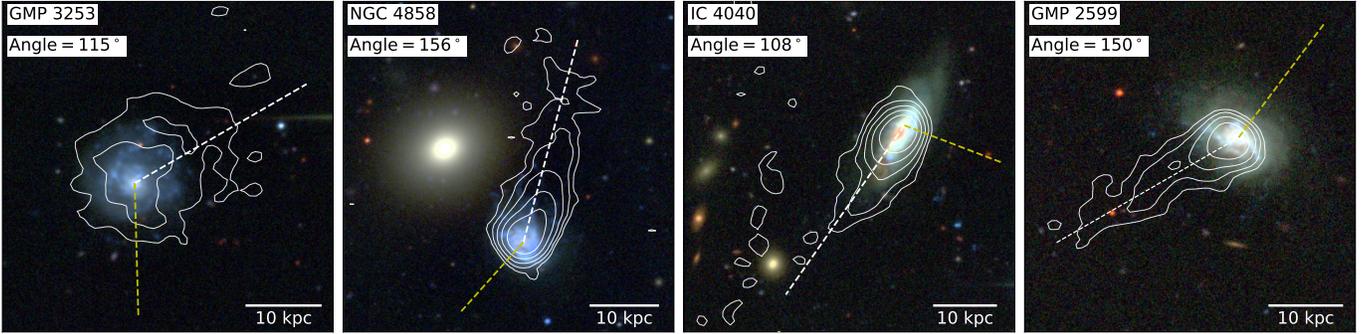

**Figure 2.** Example images of four Coma cluster galaxies from R21a. Similar images of all of the galaxies in our sample can be seen in the Appendices in R21ab. The background RGB images consist of the i-, g-, and u-bands from the CFHT and the contours are LOFAR 144 MHz continuum. The lines indicate the measured tail direction (white dash) and the direction to their host's center (yellow dash), and the angle between them is shown beneath the galaxy title in the upper-left corner.

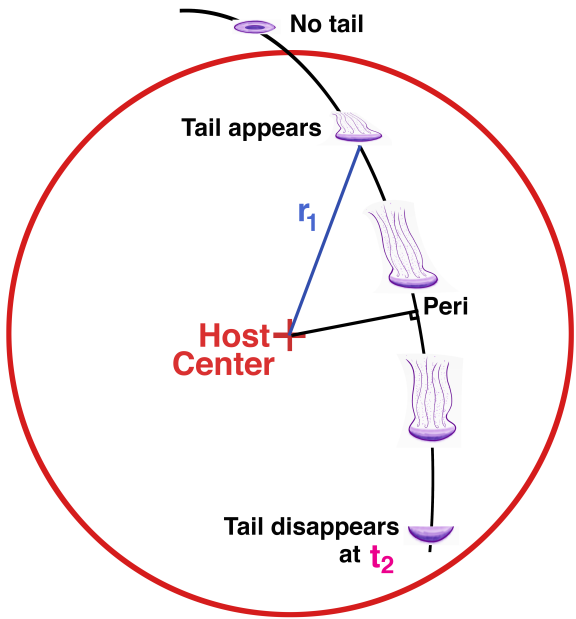

**Figure 3.** Cartoon schematic illustrating the key parameters of the model. For satellite halos on first infall into the host halo, tails first appear at a specified distance $r1$ from the host center, provided as a percentage of the host halo's $r_{200}$ (indicated by the red circle). The tails generally point in the opposite direction to the orbital trajectory of the satellite halo through the host halo as shown in the schematic, but the delay parameter $\delta$ (in Myr) allows the tail direction to have a delayed response if the orbital trajectory changes direction rapidly. The tails shown in the schematic point exactly opposite to the orbital trajectory line meaning $\delta$=0 Myr. Finally, the tail disappears from view a specified time $t2$ (in Myr) after the pericenter occurs, where $t2$ can be negative if the tail disappears before pericenter.

$\delta$: In the simplest scenario, we assume the tail direction points back in the opposite direction to the motion of the halo with respect to the host's frame of reference. However, it is known from hydrodynamical simulations that gas being stripped from a galaxy does not respond instantly to ram pressure and takes a finite time to be accelerated (Roediger & Brüggen 2007). Similarly, when galaxies rapidly change their orbital direction (for example, near pericenter), a finite time will be required for the tail to change direction, which is expected to be of order the time to form a new tail by travelling from the disk out to the end of the tail. Therefore, we introduce the delay parameter $\delta$ which is the delay (in Myr) of the response of the tail direction to a change in orbital direction. We allow $\delta$ to vary from 0 to 500 Myr in 100 Myr timesteps. For example, $\delta$ = 0 Myr corresponds to the tail direction pointing directly opposite to the orbital motion of the halo with respect to the host measured at that instant, meaning there is no delay. Meanwhile, if $\delta$ = 200 Myr, the tail direction points opposite to the orbital motion measured 200 Myr earlier, meaning the change in tail direction has been delayed by 200 Myr.

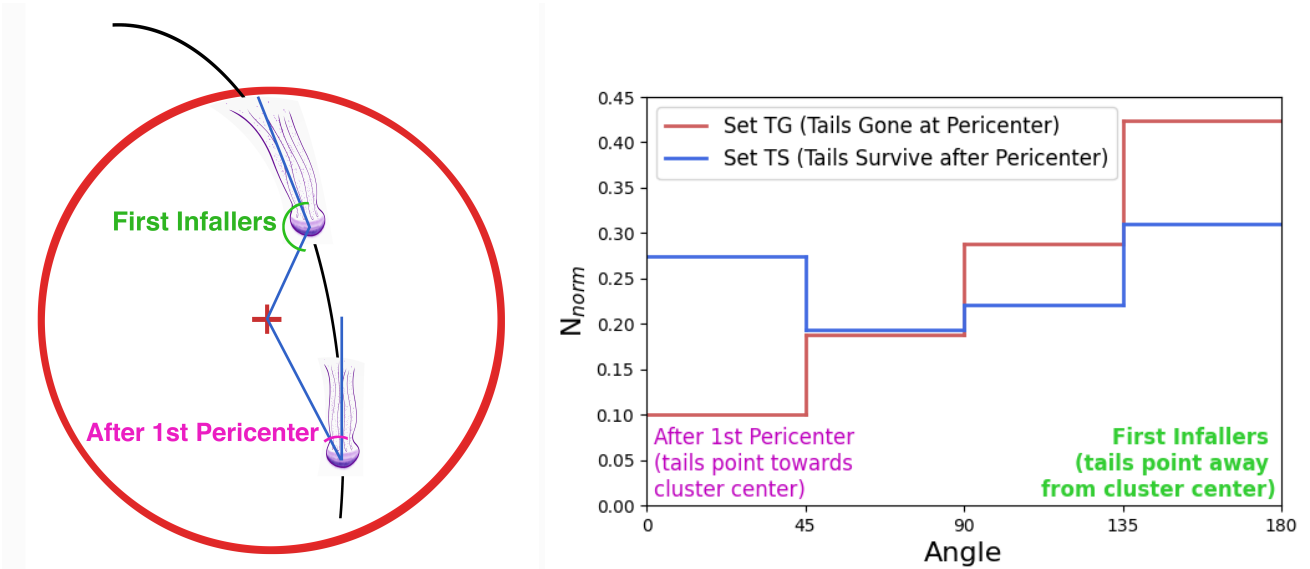

**Figure 4.** Left: Cartoon illustrating how the 3D angle of the Jellyfish galaxy tail (measured with respect to the host center) depends on whether the galaxy is on first infall (large angle shown with green arc) or has passed orbital pericenter (small angle shown with pink arc). The red circle illustrates the cluster $r_{200}$. Right: For illustrative purposes, we present the projected tail angle histograms of two sets of model galaxies. In the first set (red histogram, labelled 'Set TG'), the galaxies lose their tails at the instant they reach pericenter ($r1$, $\delta$, $t2$=100, 500, 0). The second set (blue histogram, labelled 'Set TS') is identical except they maintain their tails for 500 Myr after pericenter ($r1$, $\delta$, $t2$=100, 500, 500). The red histogram illustrates that projection effects alone can allow some galaxies to fall into the lowest angle bin (0-45°), but an increase in the object number in this bin (as we will later show is found in our observed sample, see top panel of Fig. 8) requires galaxies to maintain their tails after pericenter passage.

The 500 Myr upper-limit of $\delta$ was estimated from hydrodynamical wind-tunnel simulations (Tonnesen 2019) based on the time required to form a new gas tail of similar length to those observed (tens of kiloparsecs). We also repeated our tests using a more extreme 1 Gyr upper-limit for $\delta$ but our results were very similar and did not alter our main conclusions.

$t2$: Finally, we record the moment when the halo passes pericenter (the minimum 3D distance from the host center along the halo's orbit), and then assume the tail remains visible for a chosen length of time ($t2$) in Myr after the moment of pericenter. $t2$ varies from -400 Myr to +2000 Myr, in 100 Myr steps. The negative value of the lower limit corresponds to the tail disappearing from view as early as 400 Myr before pericenter, while positive values correspond to tails disappearing after pericenter. As with $r1$, we chose a generous range of $t2$ to allow a wide area of parameter space to be explored.

By 'painting on' the tails in this manner, we are able to quickly and efficiently cover a large area of parameter space, varying all three parameters in the steps described above to form a 3D grid of the models, where each grid point represents a unique combination of $r1$, $\delta$ and $t2$ values. In principle, we could paint the tails onto the simulated halos at a chosen instant (e.g., z=0) to mimic the redshift of the observed galaxies. However, in practice we find the measured tail angle distributions change very little over the last several Gigayears, because the orbits of the halo population do not change significantly on such short timescales. Therefore, we can stack the results over the last two Gigayears (20 snapshots), which greatly increases our statistics without altering the shape of the tail angle distributions. As a result of this and the large numbers of simulated hosts and first infaller satellites, there are excellent statistics contained in almost every 3D grid point of the model in $r1$, $\delta$ and $t2$ space. 98% of the grid points contain samples of greater than 1000 tails. Only 0.02% of the grid points contains a sample of less than 106 tails (the size of the observed sample). For each set of parameters (e.g., $r1$, $\delta$, $t2$), the tails are painted on in 3D, and we then project them down a single line of sight to produce their projected (2D) tail angle distributions, to match the observed projected tail direction histograms. As a result of good statistics in the number of hosts and first infaller satellites, we find negligible changes in the histograms for different choices of line-of-sight. We also see little evidence for a clear change in the shape of the histograms if we split the infalling halos by mass. Similarly, we also take their 3D velocities and radii within the host halos and project them into a projected (2D) phase-space diagram (line-of-sight velocity versus pro-jected radius from the host halo center), and count the relative numbers in each region of phase-space, as we do for the observed JF as described in Sect. 3.3.

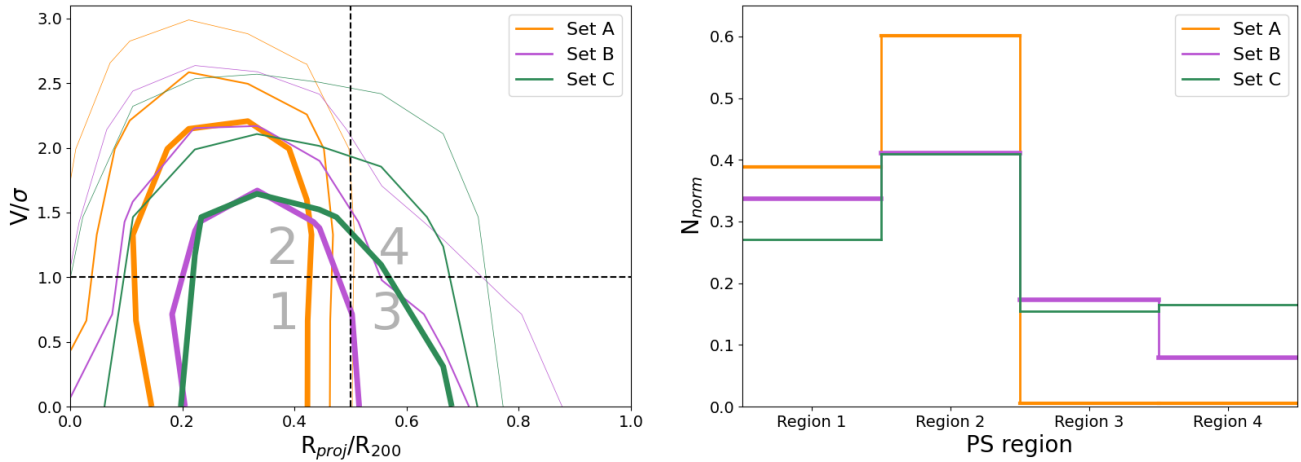

**Figure 5.** The left panel illustrates the four regions in phase-space that we consider (labelled '1-4'). For illustrative purposes, we consider three sets of model galaxies. The model parameters of Set A, B and C are ($r1$, $\delta$, $t2$)=(50, 500, 0), (50, 500, 1000), and (75, 500, 0) for orange, purple, and green, respectively. Set A and Set B are identical except tails last for 1000 Myr after pericenter in set A while they disappear at pericenter in set B. Set B and Set C are identical except the tails first become visible at 50% of the host's $r_{200}$ in set B compared to 75% for set C. The thick, medium and thin contour in the left panel contains 25, 50 and 75% of the set, respectively, and right panel shows their histograms. See text for further details of the three model data sets.

### 3.2. *Projected Tail Directions*

For each observed JF galaxy, we visually estimate the projected angle of the tail with respect to the host's center, as described in R21ab. A tail pointing directly towards the host's center is defined as having a tail angle of $0°$ and a tail pointing directly away from the host center has a tail angle of $180°$. We build up a histogram of the tail angles, using 45 degree wide angle bins. As the angle bins are chosen to be quite wide, the overall shape of our histograms are not sensitive to small inaccuracies in the measurement of the tail angle.

The left panel of Fig. 4 presents a schematic of how the direction of a ram pressure stripped tail depends on whether it is before or after pericenter in its orbit. Objects on first infall will have high tail angles in 3D, as their tails point away from the host's center in 3D. After pericenter, objects that still have tails will have low tail angles in 3D, as their tails point towards the host's center.

However, the observed tail angles are seen in projection, not in 3D. To see how the projected tail angle histograms depend on whether galaxies keep their tails past pericenter, for illustrative purposes we compare two model data sets in the right panel of Fig. 4. The model parameters are ($r1$, $\delta$, $t2$)=(100, 500, 0) and (100, 500, 500) for the red and blue line, respectively. In the first set (red line, labelled 'Set TG'), the tails disappear at the moment of pericenter, and so in 3D all the tails point away from the host's center. But thanks to projection effects, some objects can still have projected tail angles of $<90°$, although the shape of the histogram always slopes down towards the lower projected angles (see the overall shape of the red histogram). In the second set (blue line, labelled 'Set TS'), their tails remain visible for 500 Myr after pericenter. Only if the tails can survive past pericenter can the 0-45$°$ bar actually be higher than the 45-90$°$ bar (as shown by the blue histogram). This is important as we will see in Section 5.1 that the main LOFAR JF sample also presents this feature.

### 3.3. *Locations in Projected Phase-space*

Besides the tail angle distributions, we can also use the locations of JF galaxies in a projected velocity-radius phase-space diagram as an additional observational constraint. To do so, we count the relative number of JF galaxies in four regions in phase-space. The JF galaxies are separated at $r/r_{200} = 0.5$ and $V/\sigma = 1.0$, as shown in the left panel of Fig. 5. The choice of two bins in projected distance and two bins in line-of-sight velocity means the existence of both a radial gradient and a velocity gradient can potentially influence the results.

The right panel of Fig. 5 shows how the shape of the corresponding projected tail angle histogram is sensitive to the cluster radius at which the tails first appear, and/or how long they last after pericenter. For illustrative purposes, we compare three model data sets (set A, B and C). The model parameters of Set A, B and C are ($r1$, $\delta$, $t2$)=(50, 500, 0), (50, 500, 1000), and (75, 500, 0) for orange, purple, and green, respectively. Set A (orange) versus set C (green) illustrates the change if the clustocentric radius when the tails first appear is varied from 50% to 75% of the host's $r_{200}$, respectively, while keeping the other parameters fixed. If the radius that the tails appear is larger (set C), more JF appear at larger projected radii, as expected (set C has more galaxies in region 3 and 4 than Set A). Set A (orange) versus set B (purple) illustrates the change if the duration that the tails are visible after pericenter is varied from 0 to 1000 Myr while keeping the other parameters fixed. If the tails can last a long time after pericenter (set B), more objects appear at larger radii because they move out to these radii after passing pericenter (set B has more galaxies in region 3 and 4, especially in region 3). As a result, by comparing the sets, we can see how varying both the radius at which the tails first appear and how long they last after pericenter alters the distribution of the galaxies within the phase-space regions (see right panel of Fig. 5). As a result, estimates of where JF tails appear and how long they remain visible are much more constrained if we also consider their distributions in phase-space alongside the tail angle measurements. This combination of measurements allows us to better separate first infallers from objects that have passed pericenter. We will directly demonstrate the individual roles of the phase-space and tail angle measurements for constraining the model in Section 4.

### 3.4. Bayesian Parameter Estimation

In order to estimate the three model parameters ($r_1$, $\delta$, and $t_2$), we adopt the Bayesian approach and sampled the posterior using the Markov chain Monte Carlo (MCMC) method (see Sharma 2017). Assuming that the observational data are from Normal distributions, we set the likelihood as below

$$\ln L = -\frac{1}{2} \sum_i \frac{(obs_i - model_i)^2}{obserr_i^2}. \qquad (2)$$

where the $obserr_i$ is provided by bootstrapping. As there are four angle bins and four phase-space region bins, tail angles and phase-space location have equal weight when matching the model to the observations.

As presented in Table 1, we use uniform priors for all three model parameters over the modeled parameter range, as we do not have a clear reason to give preference to any particular values, at least within the specified ranges. These priors guarantee the posterior propriety emphasized in Tak et al. (2018); that is, the resulting posterior is a probability distribution.

**Table 1.** Adopted Priors

| parameter | prior type |
|---|---|
| $r_1/r_{200}$ | uniform over (10, 200) in % |
| $\delta$ | uniform over (0, 500) in Myr |
| $t_2$ | uniform over (-400, 2000) in Myr |

To sample the posterior, which is proportional to the product of likelihood and prior, we employ the affine-invariant ensemble sampler called emcee (Foreman-Mackey et al. 2013, 2019). We sample the posterior as in Shinn (2020), monitoring the convergence of the MCMC sampling with the integrated autocorrelation time ($\tau_{int}$) which is defined as below.

$$\tau_{int} = \sum_{t=-\infty}^{\infty} \rho_{xx}(t), \text{ where } \rho_{xx}(t) = \frac{\mathbb{E}[(x_i - \bar{x})(x_{i+t} - \bar{x})]}{\mathbb{E}[(x_i - \bar{x})^2]}. \qquad (3)$$

Here $\rho_{xx}$ is the autocorrelation function for the sample sequence $\{x_i\}$, $t$ is the time difference—or distance—between two points in the sequence $\{x_i\}$, $\bar{x}$ is the mean of sequence $\{x_i\}$, and $\mathbb{E}[\cdot]$ means the expectation value.

When the $\tau_{int}$ values of all the three parameters are low enough at the final iteration to have the effective sample size (ESS) of $> 2000$, we stop the sampling (see Shinn 2020, for the detail). This condition means that we have $> 2000$ independent samples for the probability distributions of the three parameters. We note that the abrupt, long-lasting increase of $\tau_{int}$, which cause the abrupt ESS decreases (worsening of the convergence), were also observed during the sampling as reported in Shinn (2020); hence, we emphasize that convergence monitoring is important during the MCMC sampling.

Although in the model we paint JF tails onto all the first infaller halos, we note that the comparison between the observed and simulated galaxies is done with normalised histograms. This means that we are not necessarily assuming that all galaxies infalling into massive hosts form tails. Rather, we assume that the observed galaxies exist in a randomly chosen subsample of the first infaller halos in the cosmological simulation.

## 4. TESTING THE METHOD WITH MOCKS

In order to test the approach we use when comparing the observations with the models, we create two mock observational data sets. These are selected from the model grid. Mock 1 and Mock 2 have ($r1$,$\delta$,$t2$) = (80, 400, 600) and (70, 200, 400), respectively. We will later see in Section 5.1 that these choices of mock fall in a similar area of parameter space as the median results of our observed JF sample. From the model, we directly have the normalised histograms of the angle bins and phase-space regions, and we additionally give each data point an error bar of similar size to those of the observational sample (in this case, we assume a fixed error of 0.05 on each normalised histogram bar).

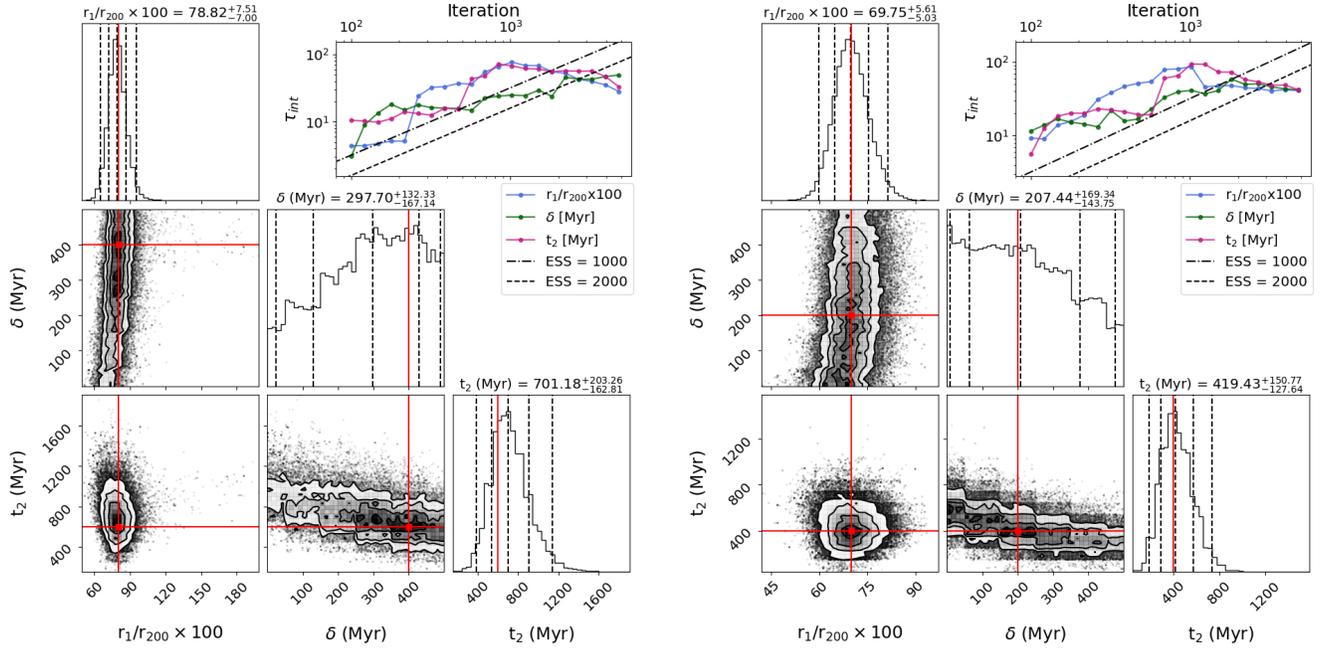

**Figure 6.** Results of the MCMC sampling for Mock 1 (left) and Mock 2 (right). For each mock test, the panels are arranged as follows. In the upper-right, a convergence monitoring panel is shown for each parameter (see legend). The other panels take the form of a corner plot. Panels with grey scale shading and contours are 2D PDFs comparing two different model parameters. The upper-left, center and lower right panel is a marginalized PDF of $r1/r200$, $\delta$, and $t2$, respectively. In the marginalized PDF panels, the central vertical dashed line is the median of the distribution while the surrounding vertical dashed lines show the 68% and 95% credible interval. The subtitles of the panels provide the median value and the errors are the 68% credible interval. The red lines shows the input value for that parameter. Our method is successful at reliably constraining the input values used to make both the mocks – for all the parameters, the input value falls within the 68% credible interval around the median of the marginalized PDFs.

The left and right column in Fig. 6 show the results for Mock 1 and Mock 2, respectively. The upper-left, central, and lower-right panel of each column are the marginalized PDFs of $r1/r_{200}$, $\delta$, and $t2$. The red lines show the values of the input parameters ($r1,\delta,t2$) for each mock. In both mocks, the PDFs of $r1/r_{200}$ (upper-left) and $t2$ (lower-right) are well behaved in that they show a clear single peak. The central dashed vertical line is the median of the PDF. The other two vertical dashed lines encompass 68% and 95% of the distribution on either side of the median. For $r1/r_{200}$ and $t2$, it is clear that the model's input value is located within the 68% credible interval for all the parameters and in both Mock 1 and Mock 2. Indeed, generally it is difficult to distinguish the median lines from the $r1/r_{200}$ and $t2$ in-put values, with the single exception of the Mock 1 $t2$ PDF. This indicates that the method provides good constraints on these two parameters. Similarly, this shows that these parameters are both influential for deciding the shape of the projected tail angle histogram and the projected phase-space region histogram.

However, in both mocks the PDF of the delay parameter $\delta$ is not so well behaved. Neither show a well defined peak, and instead we see a sloping distribution across the full parameter range, with a preference for high values of $\delta$ in Mock 1, and low values in Mock 2. Despite this, the input value still falls within the 68% credible interval. Although the relative uncertainty (defined as the width of the 68% credible interval divided by the median value) is large compared to the other parameters because of the shape of the PDF (e.g., the relative uncertainty for $r1/r_{200}$, $\delta$ and $t2$ is 0.18, 1.0, 0.52 for Mock 1 and 0.15, 1.50, 0.66 for Mock 2). This means the median $\delta$ value is more poorly constrained, which likely indicates that the delay parameter is a less important parameter for controlling the shapes of the angle bins and phase-space region histograms, at least for the tested region of parameter space.

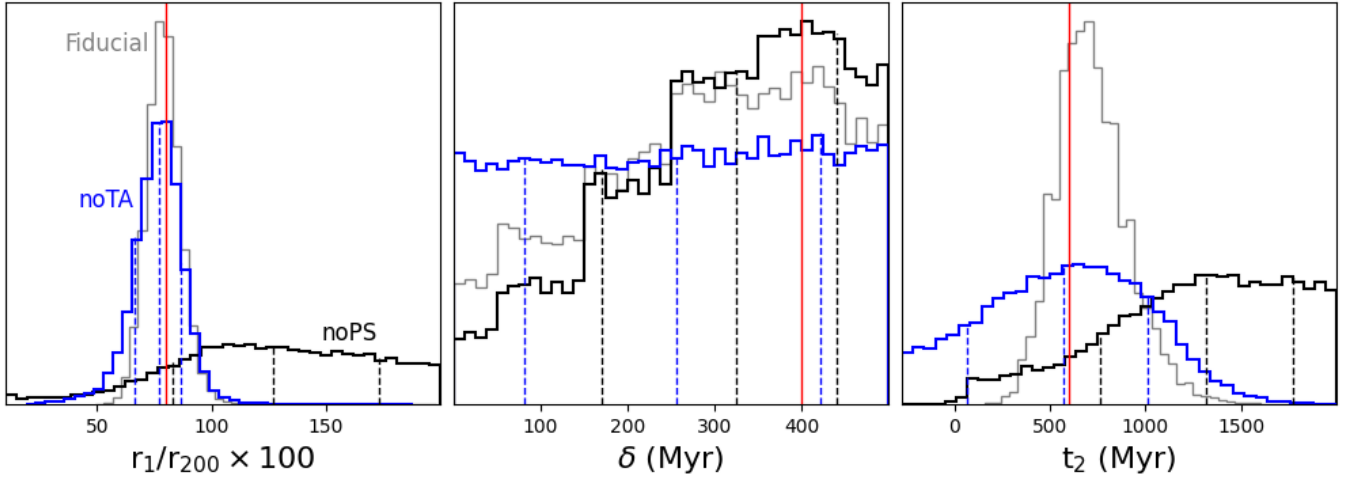

**Figure 7.** Results of the MCMC sampling for Mock 1. The red vertical line shows the input values of Mock 1. Grey histograms show the case when both tail angles and phase-space constraints are applied (labelled 'Fiducial'; identical to the results in Fig. 6). Black histograms are the same as the Fiducial case but when the constraints from phase-space are neglected (labelled 'noPS'), and blue histograms are when the constraints from tail angles are neglected (labelled 'noTA'). Vertical dashed lines show the median, and 68% credible interval (the median is the central dashed line) of each histogram. Without phase-space constraints, the Mock 1 input value fails to fall within the 68% credible limit of the 'noPS' PDF for both $r1/r_{200}$ and $t2$. Without tail angle constraints, the $t2$ distribution is much wider, meaning it is more poorly constrained.

To understand the importance of considering the locations of JF galaxies in projected phase-space in addition to their projected tail angles, we repeat the Mock 1 test but using only their projected tail angles to constrain the model. The impact this has on the marginalized PDFs of the three parameters can be seen in Fig. 7. The black histograms are the no phase-space results (labelled 'noPS'), for comparison with the grey histograms (labelled 'Fiducial') that are constrained by both tail angles and phase-space results. The PDFs of $r1/r_{200}$ and $t2$ become much broader and with a less clearly defined main peak, and with long tails towards higher values. As a result, for both of these parameters, the median value (central dashed vertical line) is no longer within one-sigma of the input value (red vertical line). However, there is a less significant change to the $\delta$ parameter PDF which already had a broad distribution even when applying the phase-space constraints.

We also consider the case where we use only the phase-space results, and neglect the tail angles, when constraining the model (blue histograms; labelled 'noTA'). In the first panel, the similarity between the grey and blue histograms shows that the tail angle measurements provide little constraint on $r1/r_{200}$ values. This is because most first infallers have quite plunging radial orbits, so the tail angle distribution should be quite similar until the galaxies approach and pass pericenter. Nevertheless, the 68% credible interval of the blue histogram is slightly wider than the grey one in the left panel

(38% wider) meaning that $r1/r_{200}$ is slightly better constrained with tail angle measurements. In the central panel, we see that the $\delta$ parameter distribution is flatter, and the credible interval is slightly wider than in the 'Fiducial' case. In the right panel, we see that the $t2$ parameter distribution is considerably wider for the 'noTA' model compared to the 'Fiducial' case (the 68% credible interval is a factor of 2.6 times wider, although the median is actually even closer to the input value). This highlights the importance of the tail angle measurements for constraining the $t2$ parameter. Finally, comparing all three histograms we see that, for Mock 1, the $r1/r_{200}$ results rely strongly on information about the galaxy locations in phase-space, while $t2$ results rely strongly on both tail angles information and locations in phase-space.

In summary, our mock tests demonstrate that, with the combination of both angle bins and phase-space region histograms, our approach can provide useful constraints on the values of $r1/r_{200}$ and $t2$. But it is difficult to tightly constrain the $\delta$ parameter in this manner, as the model is less sensitive to changes in $\delta$. With this in mind, we now attempt to apply our new methodology to the observed data set.

## 5. RESULTS AND DISCUSSION

### 5.1. *Result from the Main LOFAR Jellyfish Sample*

In this section, we begin by applying our methodology to the complete LOFAR JF galaxy sample which con-sists of 106 galaxies shared across 68 hosts. We then split the main sample into separate subsamples in the next section. In Fig. 8, we present normalised histograms of the angle-bins (top panel) and phase-space regions (bottom panel). Tail angle measurements are the same as those made in R21ab, and are defined with respect to the stellar mass weighted host center. In principle, X-ray centers might provide a better measurement of the potential minimum within a host, but consistent X-ray center measurements are not available for all of our hosts. Uncertainties (shown as error bars on histogram bars) are calculated by bootstrapping 100,000 times.

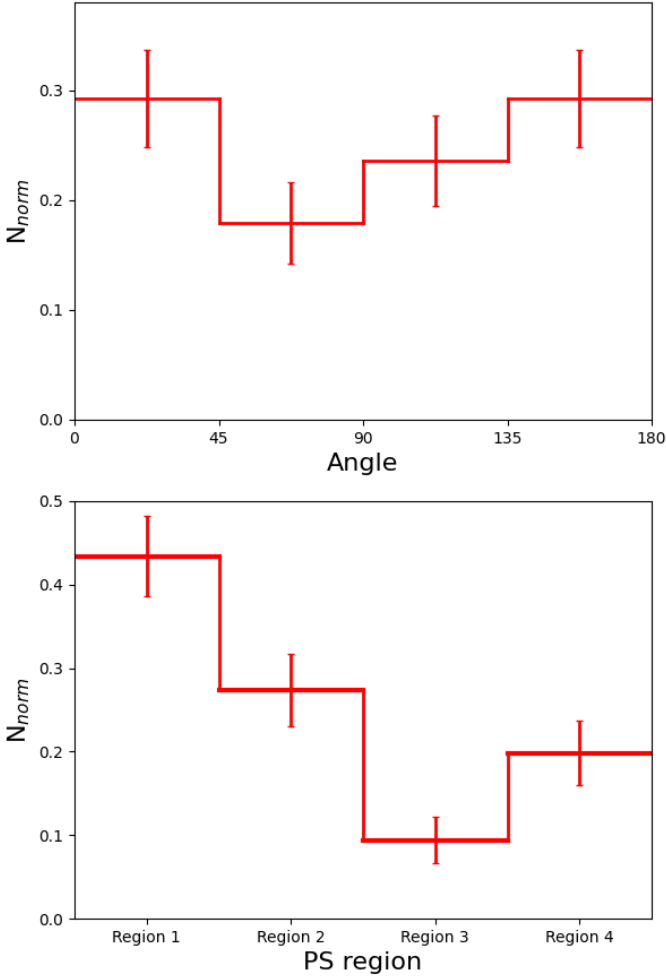

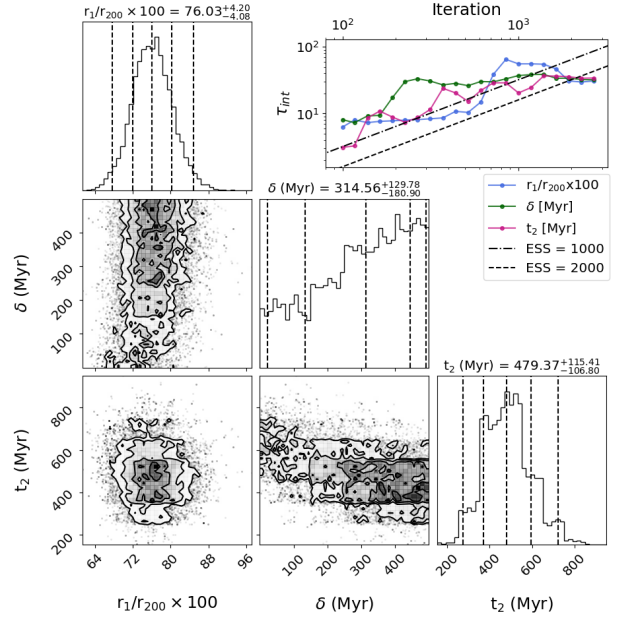

**Figure 9.** Results of the MCMC sampling for the main LOFAR Jellyfish sample. Panels are arranged as described in the Fig. 6 caption. In summary, for the main LOFAR sample, typically tails initially appear when the first infaller galaxies reach about three quarters of the host's $r_{200}$, and they last several hundred Myr after the galaxy passes orbital pericenter. The poorer constraints on the delay parameter indicate the results are less sensitive to this parameter.

**Figure 8.** Tail angle and phase-space (PS) region distributions of the main sample of LOFAR Jellyfish. Top panel: the tail angle distribution (the angle between the Jellyfish tail and host center). Bottom panel: distribution of the Jellyfish galaxies into the four PS regions we consider (the regions are illustrated in Fig. 5). Uncertainties based on bootstrapping are shown as error bars.

We note that tail-angle histogram (top panel) is peaked at high tail angles (135-180° bars) and low an-gle bins (0-45° bar) with a minimum in the 45-90° bar. As noted in Section 3, this feature already highlights that a significant fraction of the objects in the observed sample must maintain their tails for some time after pericenter passage. Meanwhile, in the phase-space region histogram, we see that most galaxies are found in region 1 and region 2 (the two regions on the left, at a projected radius $r < 0.5r_{200}$, see Fig. 5).

These histograms and their uncertainties are now used as an input to our method. The results are shown in Fig. 9. Similar to the mock tests in Section 4, the marginalized PDFs of $r1/r_{200}$ (upper-left panel) and $t2$ (lower-right panel) show a well defined single peak, while the $\delta$ parameter shows no clear peak and is less well constrained. The median value of $r1$ is $(76.0^{+4.2}_{-4.1})\%$ of $r_{200}$, and $t2$ is $(479.4^{+115.4}_{-106.8})$ Myr, while the median $\delta$ value is $(314.6^{+129.8}_{-180.9})$ Myr. Here, the upper and lower limits encompass the 68% credible interval. The relative uncertainty (defined as the width of the 68% credible interval divided by the median) for $r1/r_{200}$, $\delta$ and $t2$ is 0.11, 0.98, 0.46, which is similar to the relative uncertainties measured in Section 4 for the two mock tests. Using the simulations, we measure that the median time for a first infaller to reach 76% of $r_{200}$ is $334^{+206}_{-153}$ Myr.

In summary, the takeaway message from modelling the full LOFAR sample of JF galaxies is that, for the ensemble of JF galaxies as a whole, typically their tails appear on first infall when the objects reach roughly three-quarters of $r_{200}$ of their host and their tails survive the first pericenter, remaining visible for a few hundred Myr after the pericenter passage. A few hundred Myr is roughly the expected timescale for 144 Mhz radio continuum to decay in a ~micro-Gauss magnetic field. However, the magnetic field strengths in the tails are quite uncertain, and also the timescales could be increased by the presence of star formation in the tails or decreased if mixing strongly dilutes the tails.

| Sample | $100 \times r_1/r_{200}$ | $\delta$ (Myr) | $t2$ (Myr) |
|---|---|---|---|
| All | $76.0^{+4.2}_{-4.1}$ | $314.6^{+129.8}_{-180.9}$ | $479.4^{+115.4}_{-106.8}$ |
| Hi Galaxy Mass | $75.7^{+6.9}_{-6.6}$ | $218.2^{+163.6}_{-144.3}$ | $618.7^{+176.3}_{-154.6}$ |
| Lo Galaxy Mass | $75.7^{+6.5}_{-5.7}$ | $363.5^{+103.0}_{-198.9}$ | $398.0^{+136.8}_{-132.2}$ |
| Hi Host Mass | $84.2^{+8.1}_{-6.7}$ | $262.9^{+163.2}_{-179.9}$ | $433.5^{+164.0}_{-147.1}$ |
| Lo Host Mass | $67.4^{+5.8}_{-5.8}$ | $275.3^{+149.6}_{-172.0}$ | $566.8^{+153.9}_{-136.6}$ |
| Hi Mass Ratio | $70.7^{+5.9}_{-5.5}$ | $250.7^{+159.2}_{-161.6}$ | $564.6^{+149.3}_{-139.0}$ |
| Lo Mass Ratio | $81.3^{+8.1}_{-6.5}$ | $300.7^{+146.4}_{-196.9}$ | $423.0^{+177.3}_{-139.8}$ |

**Table 2.** Results of the MCMC sampling for the various subsamples of the main LOFAR Jellyfish sample (as illustrated visually in Figure 10). The values given in each cell of the table are the median and the 68% credible interval values of the marginalised PDF for each subsample (format: median ± credible interval).

## 5.2. Subsamples of the Main LOFAR Jellyfish Sample

The main LOFAR Jellyfish sample is a mixture of different mass hosts and different mass satellites (see Fig. 1). More massive hosts might be expected to submit their satellites to more powerful ram pressures through a denser ICM and faster orbital velocities. Similarly, lower mass satellites are expected to be more sensitive to those environmental effects (Gullieuszik et al. 2020). Therefore, in the following, we split our main sample in half into two equal sized subsamples[1] according to host mass (split at $10^{14}$ M$_\odot$), galaxy stellar mass (split at $10^{10.2}$ M$_\odot$), and the ratio of the satellite-to-host mass (split at a ratio of $1.5 \times 10^{-4}$). The resulting distributions of tail angles and location in phase-space are shown in Appendix C (Fig. 13).

We then carry out our analysis on each subsample individually. The PDFs for all three parameters are shown as violin plots in Fig. 10. The horizontal dashed line shows the median and the dotted lines encompass the 68% credible interval (the full set of median and 68% credible interval values are additionally provided in Table 2). For comparison purposes, the results for the main sample are shown first on the left. Besides comparing the PDFs by eye, we also sample from the PDFs one million times, and compute the percentage probability that the value drawn from the PDF of the 'Hi' subsample is greater than a value drawn from PDF of the 'Lo' subsample ($P_{\mathrm{Hi>Lo}}$, values shown in lower-left corner of each subsample panel). In this way, for identical distributions we expect $P_{\mathrm{Hi>Lo}}$=50%, if $P_{\mathrm{Hi>Lo}} \gg 50\%$ then the 'Hi' subsample PDF strongly prefers higher values compared to the 'Lo' subsample, and if $P_{\mathrm{Hi>Lo}} \ll 50\%$ then the 'Hi' subsample PDF strongly prefers lower values than the 'Lo' subsample. We confirm that the $P_{\mathrm{Hi>Lo}}$ values shown do not change if we were to sample the PDFs in excess of one million times.

[1] We also repeated this test using a more extreme upper third vs lower third percentile and found qualitatively consistent, albeit more noisy, results (due to reduced statistics).

Moving from left to right, we consider the various subsamples in turn.

*Galaxy Mass:* $r_1/r_{200}$ shows negligible change between the subsamples ($P_{\mathrm{Hi>Lo}} \approx 50\%$). But, the median $t2$ is larger for the high mass satellites (618 Myr) than for the low mass satellites (398 Myr), and they differ by greater than their 68% credible intervals, with $P_{\mathrm{Hi>Lo}} = 70.6\%$. Physically, this means that the moment at which the tails start to become visible is not sensitive to galaxy mass, but more massive galaxies maintain their tails for longer after pericenter passage. This could perhaps be the result of galaxies having similar self-gravity in the outskirts of their disks when tails first form. But stronger self-gravity near the centers of massive disks could perhaps enable them to hold onto their gas for longer (Jaffé et al. 2018). Or, the presence of larger quantities of gas overall in more massive disks could also play a role in extending their visibility after pericenter as they do not need to lose such large fractions of their disk gas to make a visible tail. We confirm that the median host mass changes negligibly between the Galaxy Mass subsamples. Lower mass galaxies tend to have a longer $\delta$ parameter (363 Myr) than higher mass galaxies (218 Myr), although the medians fall inside of each other's 68% credible interval, and $P_{\mathrm{Hi>Lo}} = 29.4\%$ which is a moderate difference compared to other subsamples.

*Host Mass:* Unlike with galaxy mass, $r_1/r_{200}$ clearly depends on host mass, with more massive hosts causing galaxy tails to appear further out from the host's center (at 84% of $r_{200}$ compared to 67% for low mass hosts). The difference between their two PDFs is the largest that we find among our subsamples, with $P_{\mathrm{Hi>Lo}} = 96.9\%$, and no overlap between their 68% credible in-tervals. Physically, we interpret this as more massive hosts causing stronger ram pressures in their outskirts, perhaps due to a denser ICM in the outskirts and/or higher infall velocities. Meanwhile, higher mass hosts also cause a lower $t2$ parameter (434 Myr, compared to 567 Myr for lower mass hosts). With $P_{\mathrm{Hi>Lo}} = 27.1\%$, the difference is of similar order to that shown between the 'Galaxy Mass' subsamples. This is consistent with the idea that high mass hosts also have stronger ram pressures near pericenter which can remove the tails more rapidly following pericenter. We confirm that the median galaxy mass changes negligibly between the subsamples. Meanwhile, the $\delta$ parameter distribution does not show significant differences between the subsamples, with $P_{\mathrm{Hi>Lo}} \approx 50\%$.

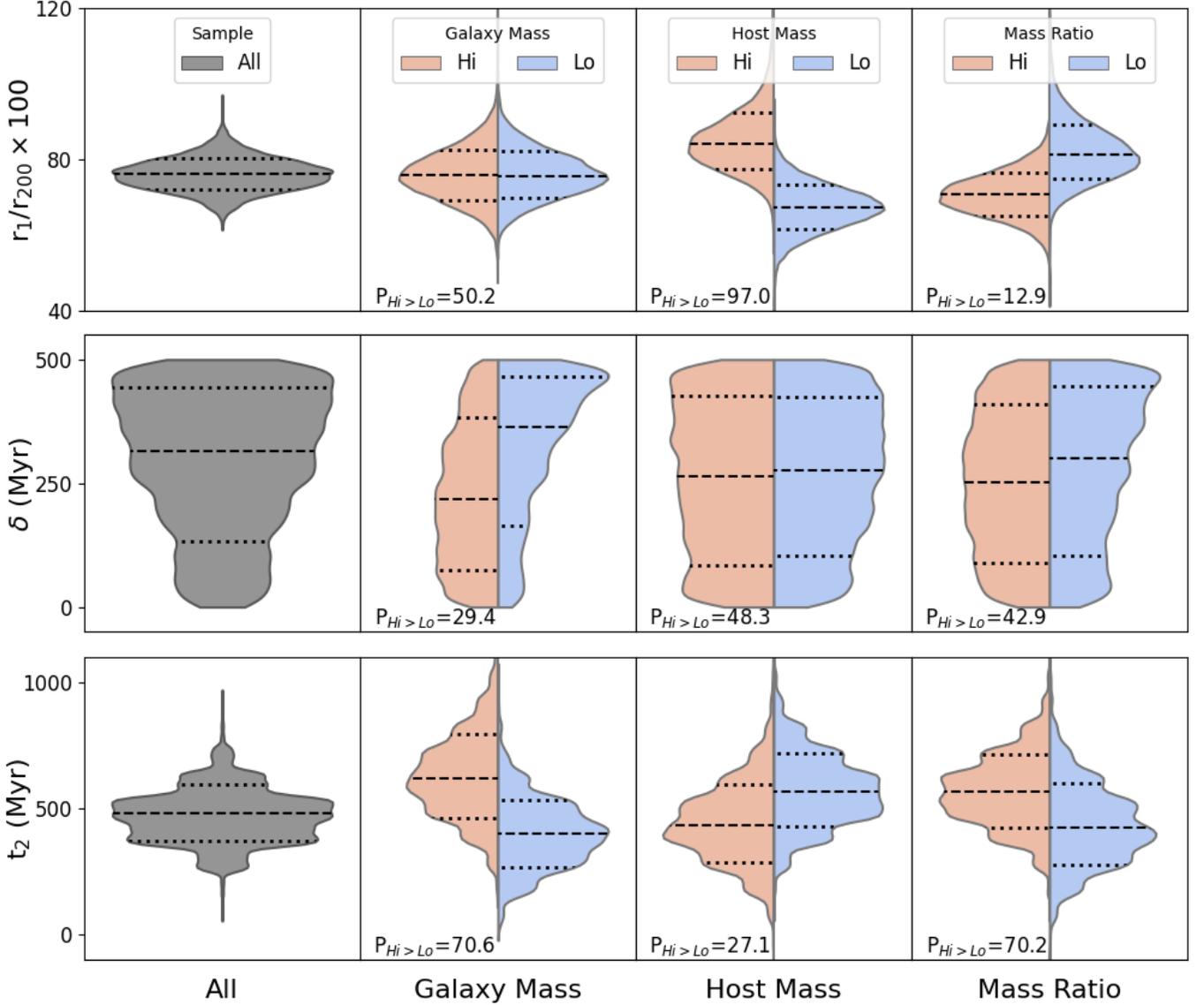

**Figure 10.** Results of the MCMC sampling for the various subsamples of the main LOFAR Jellyfish sample that we consider. The PDFs are shown as violin plots, where the horizontal dashed line is the median and the surrounding dotted lines encompass the 68% credible interval. On the far left, we show a single violin plot for the total sample in grey. Then, we show violin plots for each subsample (from left to right; galaxy stellar mass, host mass, and satellite to host mass ratio), where violin colour indicates if the subsample was the upper or lower half of the sample. Higher host masses and smaller mass ratios cause the tails of first infaller galaxies to initially appear further out from the host's center. Lower galaxy masses, lower mass ratios, or higher host masses cause tails to disappear more quickly after pericenter. The $P_{\mathrm{Hi>Lo}}$ value (shown in the lower-left corner of each subsample panel) is the percentage probability that the parameter for the 'Hi' subsample is greater than the one for the 'Lo' subsample (see text for further details).

*Mass Ratio:* As with Host Mass, both $r1/r_{200}$ and $t2$ show a dependency on the mass ratio. Lower mass ratios cause tails to appear farther out (the median $r1/r_{200}$ is 81% compared to 71% for the higher mass ratio, differing by more than the 68% credible limit) and to disappear earlier after pericenter ($t2$ is 423 Myr compared to 565 Myr for the higher mass ratio). The dependency of $r1/r_{200}$ on mass ratio is strong with $P_{\mathrm{Hi>Lo}} = 13.0\%$. For comparison, $t2$ has $P_{\mathrm{Hi>Lo}} = 70.3\%$, which is of similar order to $P_{\mathrm{Hi>Lo}}$ of the Galaxy and Host Mass subsamples. A lower mass ratio would be expected to cause the galaxy to be more sensitive to ram pressure as either its stellar mass is lower or it is in a more massive hosts (or both), and hence the appearance of the tail further out, and their disappearance more quickly after pericenter does make physical sense. As with Host Mass, the $\delta$ parameter distribution does not show large differences between the Mass Ratio subsamples ($P_{\mathrm{Hi>Lo}} = 42.9\%$).

### 5.3. Caveats

One caveat of our modelling is we assume that only the orbit through the main host halo can dictate the visibility and direction of the JF tails. This neglects the possibility that some of the observed JF tails could have formed through interactions with the hot gaseous halos of substructures in the outskirts of hosts. We do not expect this to be a major issue as the amount of contamination of this type must be generally quite low - our sample contains many group-mass hosts whose substructure must be even lower mass, and low-mass groups of galaxies are known to have very low JF fractions ($< 5\%$, R21b). However, the contamination may be higher in more massive hosts, especially in the case of very irregular hosts, but even in this case we can expect that their tails would tend to be quite randomly orientated with respect to the main host's center. Therefore, when multiple hosts are combined together, they would be unlikely to bias the overall shape of the projected tail angle distribution. We also confirmed that excluding infaller halos that enter the main host as the satellite of another more massive halo had a negligible effect on their projected tail angle distributions.

Another caveat is that some of our observed JF tails may be the result of tidal interactions with a nearby companion galaxy. Best efforts were made to exclude obvious cases of tidally interacting galaxies in R21ab, but it is not possible to fully rule out this possibility. Nevertheless, we would expect that such objects would have tail directions that are primarily a function of the orbital interactions with their neighbours and thus quite randomly orientated with respect to the host center. Thus, as with substructure, their presence would be unlikely to alter the overall shape of the tail angle distribution.

An alternative issue is that some galaxies that undergo ram pressure suffer unwinding of their spiral structure

(Bellhouse et al. 2021), and these morphological features are frequently observed in populations of blue cluster galaxies (Vulcani et al. 2022). In this case, simulations have revealed that their tails do not always point exactly downwind of their disks if the stripped material was previously rotating rapidly, and especially if their disks are viewed face-on. However, as we use wide angle bins (width of 45°) and because the observed galaxy disks will be viewed with a range of inclinations, we can expect this to be a mild source of noise in the tail angle distribution but without significantly modifying its overall shape.

We emphasise that the median values of $r1$, $\delta$, and $t2$ that we recover for a galaxy sample should be considered typical values for the ensemble of galaxies as a whole, but individual galaxies in the sample may differ from these values considerably according to other parameters that we have so far marginalised over (e.g., density of the intracluster medium or dynamical state of the host). It would be interesting to test additional parameters in the future, beyond those we considered so far (host mass, galaxy mass and mass ratio). It may be difficult to apply our method to clusters that are actively undergoing major mergers, where the cluster center is poorly defined and galaxy orbits and dynamics of the intracluster medium may be highly perturbed (McPartland et al. 2016; Roman-Oliveira et al. 2019). We also add that the reliability of our results will depend on whether JF galaxies actually do follow the model's prescription of a tail appearing at $r1$ and then remaining visible until $t2$, as currently assumed.

Finally, we note that our choice for the boundaries between the phase-space regions was fairly arbitrary, and it may be possible to improve the constraints even further by adjusting the locations of the regions in future works. Nevertheless, our tests with the mocks (e.g., see Fig. 7 and discussions in the main text) reveal that the chosen regions already provide valuable constraints.

## 6. CONCLUSIONS

In this study, we present a new and original method for constraining where the tails of observed Jellyfish (JF) galaxies appear and how long they last following their orbital pericenter. We demonstrate our approach on a sample of 106 JF galaxies identified in the LOFAR two meter sky survey (LoTSS), that exist inside 68 hosts such as groups and clusters. In star-forming galaxies, LoTSS is sensitive to synchrotron emission from cosmic rays accelerated by supernovae. For star-forming galaxies experiencing ram pressure stripping, tails of synchrotron emission may be observed as these cosmic rays are stripped from their star-forming galactic disks.

We measure the distribution of their tail angles with respect to the stellar mass weighted host centers. We also measure their distribution into regions of a phase-space plot of host-centric line-of-sight velocity vs projected radius. We then compare these two observed distributions (tail angles and phase-space regions) to predictions from models based on cosmological simulations. The novelty of our approach is to 'paint-on' the JF tails onto the dark matter halos of our N-body cosmological simulations, using a three parameter model ($r1$, $\delta$, $t2$) to decide at which radius the tails first become visible ($r1$ as a percentage of the host's $r_{200}$), how long the tails are visible after pericenter passage ($t2$ in Myr), and a delay parameter ($\delta$ in Myr) that accounts for the delay in tail directions to respond to rapid changes in orbital direction. In this way, we can quickly cover a large area of parameter space, and produce distributions of the projected tail angles and locations in the phase-space regions for comparison with the observations. The model is also highly flexible and thus the tail properties are not limited by treatments of hydrodynamical processes and sub-grid physics that could artificially influence our results. Based on a Bayesian parameter estimation using the Markov chain Monte Carlo (MCMC) method, we obtain the PDFs of the parameters ($r1$, $\delta$, $t2$) for our observed sample.

Our main results can be summarised as follows:

- Tests with mock observational data sets that have similar uncertainties as in our observed sample demonstrate that our method can provide useful constraints on both $r1$ and $t2$, when the distributions of both the angle bins and phase-space regions are combined. $\delta$ is not found to be an influential parameter for controlling the shape of the distributions.

- Applying the method to the full sample of 106 LOFAR JF galaxies, the Bayesian parameter estimation returns PDFs with median values of $r1 = (76.0^{+4.2}_{-4.1})\%$, $t2 = (479.4^{+115.4}_{-106.8})$ Myr, and $\delta = (314.6^{+129.8}_{-180.9})$ Myr, where the limits encompass the 68% credible interval. Thus, typically tails appear shortly after the satellites have crossed the host's $r_{200}$ for the first time, and the tails survive the first pericenter passage and are still visible several hundred Myr later.

- We split the full sample of LOFAR JF in half into two equal sized subsamples according to satellite mass, host mass, and the mass ratio of the satellite to the host. In general, we find that satellites which would be expected to be more influenced by ram pressure (e.g., those in more massive hosts or with lower masses with respect to their host) tend to present tails further out in the host and also lose their tails earlier after pericenter.

This study demonstrates that our novel approach to modelling the observed tail directions of JF and their locations in phase-space can provide useful physical constraints on where JF tails first appear, and highlight the interesting result that typically JF galaxies display visible radio continuum tails even after having passed pericenter. The implications of this are interesting, as we can assume that the peak strength of the ram pressure occurs at the time of pericenter. Therefore the presence of visible tails at later times could imply that gas which was stripped near pericenter has not yet fully separated from the galaxy. As the ram pressure strength may rapidly drop following pericenter, in principle some of this gas might even fall back onto their disks.

Our sample was visually selected by the presence of their JF tails, as seen in the radio continuum and, as such, the parameter estimation can be considered valid for our sample. However, this parameter estimation may not be valid for a more general sample of infalling late-type galaxies because there will likely be additional parameters (for example, gas fraction) that decide whether they form tails like those visible in our sample. In the near future, we will apply this method to alternative data sets, including samples of JF tails identified at other wavelengths such as in optical (e.g., Poggianti et al. 2016a; Roberts & Parker 2020b, Hyper-suprime cam imaging, Galaxy Zoo), HI (e.g., VIVA, Meerkat, Wallaby), Ultraviolet (e.g., Galex, UVIT), or H-alpha (e.g., VESTIGE, GASP, INT) imaging. Different components of a galaxy are expected to respond in different ways to ram pressure. For example, the atomic gas component as traced by 21cm may be stripped earliest as it is more extended, but the tails may quickly disappear if they are ionised. Other tracers may only be seen if there is recent star formation in the tails (e.g., UV). H-alpha can arise either from star formation or other mechanisms, such as shocks or ICM-ISM interactions (Poggianti et al. 2019c; Campitiello et al. 2021). Therefore, the time when tails appear and how long they last could be sensitive to the tracer that is considered. We expect that the application of our novel method to alternative data sets could provide new insights into the formation and survival of JF tails, and the physical processes occurring within them.

We thank the anonymous referee for a thoughtful report that helped improve the manuscript. BMP, MG, AM, BV acknowledge funding from the European Research Council (ERC) under the European Union's Horizon 2020 research and innovation programme (grant agreement No. 833824). BMP, MG, AM, BV also acknowledge financial contribution from the INAF main-stream funding programme (PI Vulcani) and from the agreement ASI-INAF n.2017-14-H.0 (PI A. Moretti). BV and MG acknowledge the grant from PRIN MIUR 2017 n.20173ML3WW_001 (PI Cimatti). YJ acknowledges support from FONDECYT Iniciación 2018 No. 11180558 and ANID BASAL project FB210003. JPC acknowledges partial support from FONDECYT through grant 3210709, and Comité Mixto ESO-Gobierno de Chile.

*Software:* GADGET-3 code (Springel et al. 2001), MUSIC (Hahn & Abel 2011), CAMB (Lewis et al. 343 2000), ROCKSTAR (Behroozi et al. 2013), CIGALE (Boquien 289 et al. 2019)

## APPENDIX

### A. NOISE TESTS

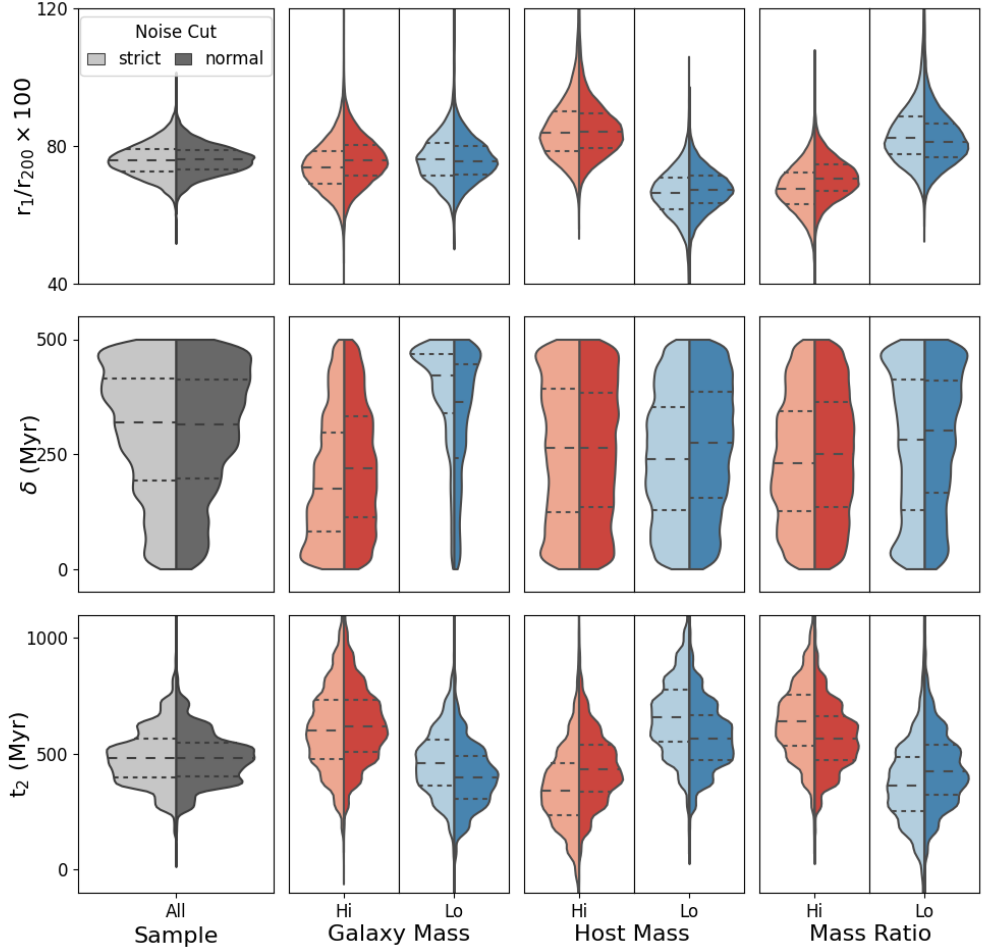

**Figure 11.** Results of the MCMC sampling for two different choices of noise cut. The marginalized PDFs of each parameter are shown as violin plots. $200\mu Jy/beam$ is our standard noise cut value (dark coloured histogram; labelled 'normal'), and is compared to the more strict value of $150\mu Jy/beam$ cut (lighter coloured histogram; labelled 'strict'). The results are overall similar and the medians (shown as horizontal long-dashed lines) agree within the 50% credible interval (shown as horizontal short-dashed lines).

As described in Section 2.1, for our main LOFAR Jellyfish sample, we choose to cut our sample for all galaxies where the background noise level is greater than $200\mu Jy/beam$. In this section, we consider how our results would change if we had chosen a more strict $150\mu Jy/beam$ cut. Fig. 11 shows violin plots of the PDFs for $r1$ (upper row), $\delta$ (middle row), and $t2$ (lower row) of the various subsamples we consider in Section 5.2. The central long-dashed line indicates the median and the surrounding short-dashed lines encompass the 50% credible interval. The darker coloured histograms are our standard noise cut of $200\mu Jy/beam$, while the more pale coloured histogram shows the more strict noise cut of $150\mu Jy/beam$. Overall, the results are similar, and all agree within the 50% credible interval. When comparing high to low galaxy mass, host mass or mass ratio subsamples, the general dependency on that parameter is conserved. This demonstrates that our main conclusions on where tails become visible and how long they last, and their general dependencies on these parameters are robust to the levels of noise in our main sample.

## B. PROPERTIES OF GALAXIES SPLIT BY TAIL DIRECTION

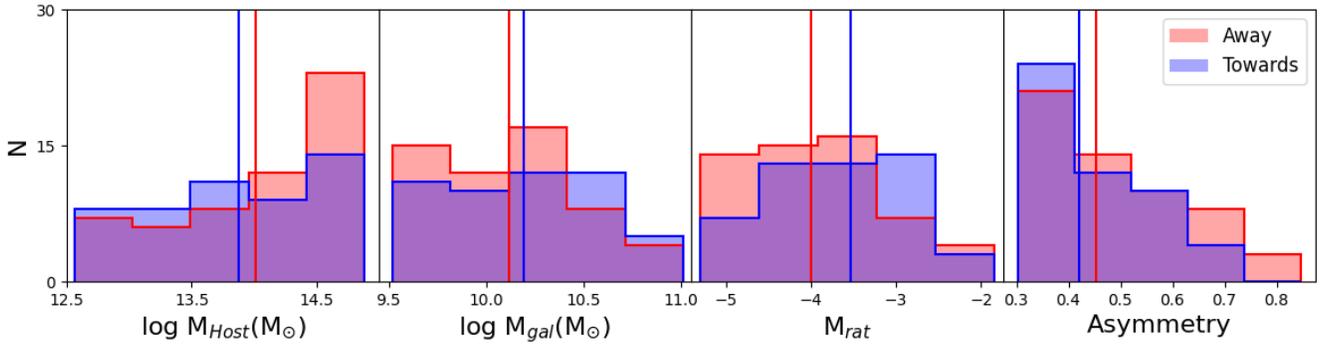

**Figure 12.** Our main LOFAR Jellyfish sample is split into two subsamples by the direction of their tail (red is pointing away and blue is pointing towards the host's center). Histograms compare the properties (host mass, galaxy mass, mass ratio, and asymmetry, shown from left to right, respectively) of the two subsamples. A vertical line indicates the median value of their distribution.

As an additional test, we split our main LOFAR Jellyfish sample into two subsamples, based on the direction of their tails with respect to their host's center. Red histograms are tails pointing away from the host's center (tail angle $>90°$, labelled 'Away', containing 50 galaxies), and blue histograms point towards it (tail angle $< 90°$, labelled 'Towards', containing 56 galaxies). We can expect that the 'Towards' subsample has more objects that have passed pericenter than the 'Away' sample (which is reasonable, as illustrated in Fig. 4).

We see from the distribution shapes and the median lines that the 'Away' sample appears to prefer higher mass hosts, lower mass galaxies, and lower mass ratios (first three panels, from left to right). In the main paper, galaxies with higher mass hosts, lower mass galaxies, and lower mass ratios tend to lose their tails more quickly (shorter $t2$) and/or form their tails earlier (larger $r1$), as shown in Fig. 10, thus they might be expected to prefer the 'Away' sample. Therefore, the results show here seem to offer support for the results in the main paper.

The fourth histogram (on the far right) shows the asymmetry distribution of the subsamples (as defined in Eqn. 1). We see that the 'Towards' subsample prefers smaller asymmetry values, which could be interpreted as them having shorter tails after passing pericenter. Physically, this could occur because the ram pressure would be expected to decrease in strength after passing pericenter. Additionally, 144 Mhz radio continuum emission is expected to fade on timescales of hundreds of Myrs, as the synchotron emitting electrons age and the emission shifts to lower frequencies, in the absence of any re-acceleration in the tail (Feretti & Giovannini 2008).

However, we add a note of caution. Although some differences between the red and blue histograms in each panel are visible by eye, according to a Kolmogorov-Smirnov (K-S) test the differences between the red and blue histograms in each panel are not very significant. If $D$ is the K-S statistic and $p$ is the corresponding $p$-value, we measure $(D, p)=(0.228, 0.13)$ for host mass, $(0.183, 0.34)$ for galaxy mass, $(0.204, 0.22)$ for mass ratio, and $(0.156, 0.48)$ for asymmetry, so all the p-values are greater than 0.05. The poor number statistics ($\sim 50$ galaxies in each subsample) likely contributes to the low significance of these results. This emphasises the value of our modeling in the main draft, which shows diverse distinctions between the various parameters (see Fig. 10) in a way that the test presented here cannot.

## C. HISTOGRAMS OF THE LOFAR JELLYFISH SUBSAMPLES

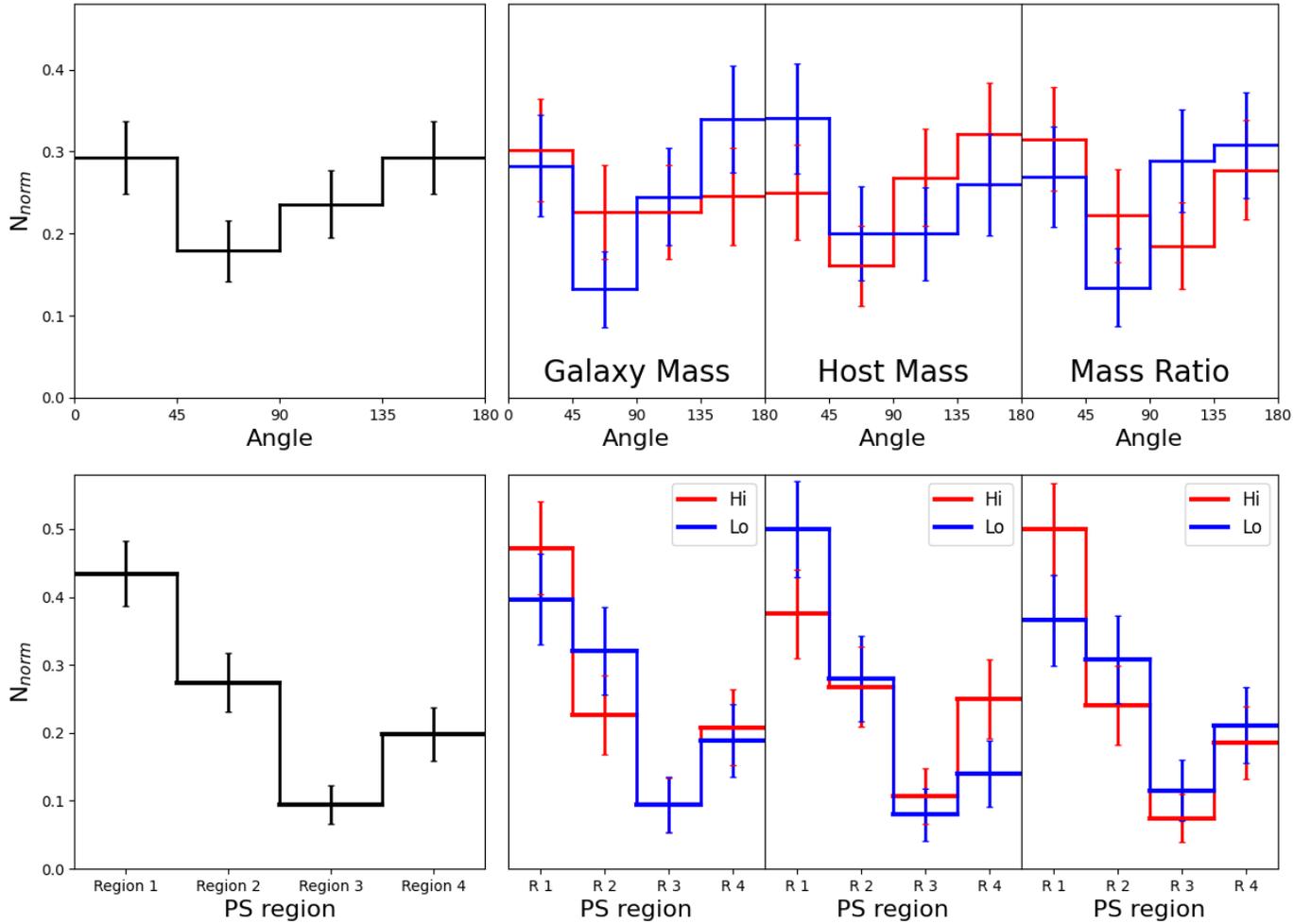

**Figure 13.** Histograms of the observed tail angle distribution (top row) and location of the samples in the phase-space regions (bottom row) for the main LOFAR Jellyfish sample (first column), and for the high (red) and low (blue) subsamples that split the main sample according to galaxy mass, host mass and mass ratio (column two to four respectively). Uncertainties based on bootstrapping are shown as error bars.

As described in Section 5.2, our main LOFAR Jellyfish subsample is split into high and low subsamples according to galaxy stellar mass (split at $10^{10.2}$ $M_\odot$), host mass (split at $10^{14}$ $M_\odot$), and the ratio of the satellite-to-host mass (split at a ratio of $1.5\times10^{-4}$). The impact on the shape of the histograms can be seen in Fig. 13. These histograms are then used as the input to our method, and the resulting PDFs of the three model parameters $r_1$, $\delta$, and $t_2$ can be seen in Fig. 10.